\documentclass[
  aps,
  prd,
  reprint,
  superscriptaddress,
  nofootinbib,
  longbibliography
]{revtex4-2}

\usepackage{amsmath}
\usepackage{amssymb}
\usepackage{bm}
\usepackage{mathtools}
\usepackage{graphicx}
\usepackage{xcolor}
\usepackage{microtype}
\usepackage{hyperref}

\newcommand{\be}{\begin{equation}}
\newcommand{\ee}{\end{equation}}
\newcommand{\bea}{\setlength\arraycolsep{2pt} \begin{eqnarray}}
\newcommand{\eea}{\end{eqnarray}}

\newcommand{\mc}{\mathcal}

\hypersetup{
  colorlinks=true,
  linkcolor=blue,
  citecolor=blue,
  urlcolor=blue,
  pdftitle={
    Axial Obstructions to Rotating Constant-Norm
    Vector Vacua
  },
  pdfauthor={Minyong Guo and Zhong-Ying Fan}
}

\newcommand{\dd}{\mathop{}\!\mathrm{d}}
\newcommand{\Lie}{\mathcal{L}}
\newcommand{\Ord}{\mathcal{O}}
\newcommand{\Kretsch}{\mathcal{K}}

\newcommand{\rplus}{r_{+}}
\newcommand{\rminus}{r_{-}}

\begin{document}

\title{
Axial Obstructions to Rotating Bumblebee Vacuum Solutions
}

\author{Minyong Guo}
\email{minyongguo@bnu.edu.cn}
\affiliation{
  School of Physics and Astronomy,
  Beijing Normal University,
  Beijing 100875, China
}
\affiliation{
  Key Laboratory of Multiscale Spin Physics
  (Beijing Normal University),
  Ministry of Education,
  Beijing 100875, China
}

\author{Zhong-Ying Fan}
\email{fanzhy@gzhu.edu.cn}
\affiliation{
  Department of Astrophysics,
  School of Physics and Materials Science,
  Guangzhou University,
  Guangzhou 510006, China
}

\date{}

\begin{abstract}
We establish two obstructions to globally regular
rotating constant-norm vector vacua. At an axial fixed point of a
regular nondegenerate bifurcate Killing horizon, horizon-boost and
axial invariance force every smooth symmetry-inheriting one-form to
vanish, contradicting a strictly nonzero constant norm.
Independently, varying axial conicity produces an orthonormal
curvature component diverging as the inverse proper distance and
precludes a \(C^2\) extension. Applied to a three-parameter
Kerr--disformal family in Einstein--bumblebee gravity, these results
reveal a nonsmooth bumblebee one-form, while direct calculation
shows that the Kretschmann scalar diverges as the inverse square of
the transverse proper distance to either open exterior axis. Even
the distinguished nonextremal branch selected by nonpolar
outer-horizon regularity is therefore an exact nonpolar rotating
exterior solution, not a globally regular isolated black hole.
Together, these field- and metric-level obstructions provide a
two-pronged no-go framework for globally regular rotating
constant-norm vector vacua.

\end{abstract}

\maketitle

\section{Introduction}
\label{sec:introduction}

Spontaneous Lorentz-symmetry breaking provides a possible
low-energy signature of physics beyond general relativity
\cite{KosteleckySamuel1989,Carroll:2001ws,
Kostelecky:1989jp,Gambini:1998it}.
Einstein--bumblebee gravity realizes this mechanism through a vector
field with a nonzero vacuum expectation value and a nonminimal
coupling to curvature
\cite{Kostelecky:1989jw,Bluhm:2004ep}.
Exact static configurations are known
\cite{CasanaEtAl2018,Chen:2025ypx}, and several Kerr-like
geometries have been proposed
\cite{DingEtAl2020,PoulisSoares2022,Ovcharenko:2026rvj}.
For a rotating solution, however, a regular metric horizon is not
sufficient: the Lorentz-breaking field and the rotation axis must
also be smooth. Metric regularity therefore need not imply
regularity of the complete metric--vector configuration.

A useful exact-solution sector is generated by a rank-one disformal
deformation of a Ricci-flat seed
\cite{PoulisSoares2022,Ovcharenko:2026rvj},
\begin{equation}
g_{\mu\nu}
=
g^{(0)}_{\mu\nu}
+
\ell\,\Omega_\mu\Omega_\nu,
\qquad
B_\mu=b\,\Omega_\mu,
\label{eq:disformal-ansatz}
\end{equation}
where \(\Omega\) is a closed one-form of constant positive norm.
This is analogous to a standard disformal transformation
\cite{Bekenstein:1992pj,Bettoni:2013diz}.
For a Kerr seed, Eq.~\eqref{eq:disformal-ansatz} yields an exact
rotating family, but the resulting one-form generically fails to
extend smoothly to the rotation axis.

A related fixed-point obstruction is known in
Einstein--\ae ther theory: a nonvanishing unit timelike aether
invariant under the horizon Killing flow cannot be smooth on the
bifurcation surface of a nondegenerate horizon
\cite{Eling:2006ec,Foster:2005fr,Adam:2021vsk}.
For a timelike aether, the horizon boost alone is sufficient.
A spacelike constant-norm vacuum may instead retain components
tangent to the bifurcation surface, so an additional isotropy action
is required to eliminate them. In a rotating geometry, such an
action occurs at the axial fixed points of the bifurcation surface.

In this Letter, we establish two general results. First, at such a
fixed point, horizon-boost and axial invariance force every smooth
symmetry-inheriting one-form to vanish. A strictly nonzero constant
norm is therefore incompatible with a smooth extension through the
horizon poles. This obstruction is local and requires neither
closedness, separability, nor the field equations. Second, if
\(\rho\) is proper distance from the axis, \(z\) proper distance
along it, and
\[
f=\sqrt{g_{\phi\phi}}
=
\alpha(z)\rho+\mathcal O(\rho^2),
\]
a Cartan-frame calculation gives
\[
R_{\hat\rho\hat\phi\hat z\hat\phi}
=
-\frac{\partial_z\ln\alpha}{\rho}
+\mathcal O(1).
\]
Thus \(\partial_z\alpha\neq0\) obstructs a \(C^2\) axial extension,
unlike a uniform conical defect.

Applying these results to a three-parameter Kerr--disformal family
in Einstein--bumblebee gravity, we find that the bumblebee one-form
violates the local axis-smoothness condition and that the conicity
varies along both open exterior axes. Direct evaluation gives
\(\mathcal K\equiv
R_{\mu\nu\rho\sigma}R^{\mu\nu\rho\sigma}\sim\rho^{-2}\), proving
that the axes are scalar-polynomial curvature singularities. A
distinguished nonextremal branch nevertheless possesses a regular
outer Killing horizon away from the poles. It is therefore an exact
nonpolar rotating exterior solution, but not a globally regular
isolated black hole of the complete metric--bumblebee system.
\section{General regularity results}
\label{sec:general-results}

We first establish two geometric results independent of the
bumblebee field equations.

\paragraph{Fixed-point obstruction.---}
Let \(k\) and \(m\) be the stationary and axial Killing fields, and
let
\(\chi=k+\Omega_Hm\) generate a nondegenerate bifurcate Killing
horizon with bifurcation surface \(\mc B\)
\cite{Wald:1984rg,Racz:1992bp,Racz:1995nh}.
At any regular axial fixed point
\(p\in\mc B\cap\mc A\),
\begin{equation}
\chi|_p=0=m|_p ,
\end{equation}
where \(\mc A\) denotes the rotation axis. For a smooth one-form
inheriting both symmetries,
\(\Lie_\chi\Omega=\Lie_m\Omega=0\), evaluation at \(p\) gives
\begin{equation}
\Omega_b\nabla_a\chi^b\big|_p=0,
\qquad
\Omega_b\nabla_am^b\big|_p=0.
\label{eq:fixed-point-invariance}
\end{equation}
The map \(\nabla_a\chi^b|_p\) generates a nondegenerate Lorentz
boost on the two-dimensional normal space to \(\mc B\), so the
first equation forces the normal projection of \(\Omega\) to
vanish. At a regular axial fixed point,
\(\nabla_am^b|_p\) generates a spatial rotation on
\(T_p\mc B\), and the second equation forces the tangential
projection to vanish. Hence
\begin{equation}
\Omega|_p=0.
\label{eq:Omega-vanishes}
\end{equation}
A smooth symmetry-inheriting one-form therefore cannot have a
strictly nonzero constant norm on any region containing such a
horizon pole. The obstruction is local and requires neither
closedness, separability, nor the field equations. An adapted-frame
proof is given in Supplemental
Material~\ref{app:fixed-point-proof}.

As a static corollary, horizon-boost invariance and spherical
isotropy force every smooth, static, and spherically symmetric
one-form to vanish on the entire bifurcation surface of a
nondegenerate horizon. A strictly nonzero constant norm is therefore
incompatible with a smooth static bifurcate extension. Neither
one-sided future horizons nor degenerate extremal horizons are
excluded by this argument.


\paragraph{Varying-conicity criterion.---}
We next determine when a nonuniform axial defect obstructs a smooth
axial extension. Let \(\rho\) be proper distance from the axis,
\(z\) proper distance along it, and
\(f=\sqrt{g_{\phi\phi}}\) the circumferential radius of the
azimuthal orbits. Away from the axis, a stationary and axisymmetric
metric can be written as
\begin{equation}
\dd s^2
=
q_{ij}\dd y^i\dd y^j
+
f^2
\left(
\dd\phi+A_i\dd y^i
\right)^2,
\qquad
y^i=(\rho,t,z),
\label{eq:axial-decomposition}
\end{equation}
where
\(q_{ij}=g_{ij}-g_{i\phi}g_{j\phi}/g_{\phi\phi}\).
Suppose that the quotient geometry has a regular \(C^2\) limit in
axial Gaussian coordinates and that
\begin{equation}
f(\rho,z)
=
\alpha(z)\rho+\mathcal O(\rho^2),
\qquad
g_{i\phi}
=
\mathcal O(\rho^2),
\label{eq:axial-expansion}
\end{equation}
with \(\alpha(z)>0\), while \(A_i=g_{i\phi}/g_{\phi\phi}\) and
the derivatives required by the curvature remain bounded.

A Cartan-frame calculation gives the universal singular term
\begin{equation}
R_{\hat\rho\hat\phi\hat z\hat\phi}
=
-\frac{\partial_z\ln\alpha}{\rho}
+
\mathcal O(1).
\label{eq:universal-varying-conicity}
\end{equation}
The bounded stationary angular mixing cannot cancel this
\(1/\rho\) behavior. Hence
\begin{equation}
\partial_z\alpha\neq0
\quad\Longrightarrow\quad
\text{no \(C^2\) extension through the axis}.
\label{eq:varying-conicity-no-C2}
\end{equation}
A regular axial segment therefore requires constant conicity.
For the standard \(2\pi\) period of \(\phi\), \(\alpha=1\)
describes an elementary-flat axis, whereas a constant
\(\alpha\neq1\) describes a uniform conical defect. A
longitudinally varying conicity instead produces an ordinary
off-axis curvature divergence. The detailed derivation is given in
Supplemental Material~\ref{app:cartan-varying-conicity}.
%


The two results are logically distinct. The fixed-point
obstruction concerns the vector vacuum and does not by itself
prove that the disformal metric has divergent curvature. The
varying-conicity criterion supplies a sufficient geometric
condition for the latter. We now apply both results to the
Kerr-disformal family solutions.\\

\section{Kerr-disformal realization}
\label{sec:Kerr-disformal}

Einstein--bumblebee gravity is described by
\begin{equation}
  S
  =
  \int\dd^4x\sqrt{-g}
  \left[
    \frac{
      R+\xi B^\mu B^\nu R_{\mu\nu}
    }{2\kappa}
    -
    \frac14B_{\mu\nu}B^{\mu\nu}
    -
    V
  \right] \,,
  \label{eq:action}
\end{equation}
where \(B_{\mu\nu}=\partial_\mu B_\nu-\partial_\nu B_\mu\).
We work in the closed vacuum sector,
\(V=V'=0\) and \(B_{\mu\nu}=0\), in which the field equations
admit the disformal map \eqref{eq:disformal-ansatz}. 
Taking Kerr as the Ricci-flat seed generally yields a separable solution to the Hamilton-Jacobi equation (see \ref{app:separable-obstruction})
\begin{equation}
  \Omega
  =
  \sqrt{\frac{r^2-q^2}{\Delta}}\,\dd r
  +
  \sqrt{a^2\cos^2\theta+q^2}\,\dd\theta \,,
  \label{eq:Kerr-oneform}
\end{equation}
where
\(\Sigma=r^2+a^2\cos^2\theta\) and
\(\Delta=r^2-2Mr+a^2\). This gives a three-parameter family stationary solution. For $q=0$, it reduces to the Kerr-like solution \cite{PoulisSoares2022}. Notice that $q$ is a separation constant, arising from the constant norm of the bumblebee field. The presence of this parameter $q$ affects global properties of the solution tremendously and makes it possible for us to find a general non-extremal rotating black hole solution, with a regular outer Killing horizon away from the polar axes.

The complete disformal metric is given by 
\begin{widetext}
\begin{align}
  \dd s^2={}&
  -\left(
    1-\frac{2Mr}{\Sigma}
  \right)\dd t^2
  -
  \frac{4Mar\sin^2\theta}{\Sigma}
  \,\dd t\,\dd\phi
  \nonumber\\
  &+
  \frac{
    \Sigma+\ell(r^2-q^2)
  }{\Delta}
  \,\dd r^2
  +
  \frac{
    2\ell
    \sqrt{
      (r^2-q^2)
      (a^2\cos^2\theta+q^2)
    }
  }{\sqrt{\Delta}}
  \,\dd r\,\dd\theta
  \nonumber\\
  &+
  \left[
    \Sigma+\ell
    \left(
      a^2\cos^2\theta+q^2
    \right)
  \right]\dd\theta^2
  +
  \frac{
    \left[
      (r^2+a^2)^2
      -
      a^2\Delta\sin^2\theta
    \right]\sin^2\theta
  }{\Sigma}
  \,\dd\phi^2 \,.
  \label{eq:full-metric}
\end{align}
\end{widetext}

\section{Horizon and off-axis regularity}
\label{sec:horizon-regularity}

We first examine the geometry away from the rotation axes. For
\(0<|a|<M\), the two Kerr roots are
\(\rplus=M+\sqrt{M^2-a^2}\) and
\(\rminus=M-\sqrt{M^2-a^2}\). However, while these surfaces are null
with respect to the inverse metric, they are not generically
regular horizons of the disformal geometry.

The Kretschmann scalar near the simple
outer root has the singular structure
\begin{align}
  \Kretsch
  ={}&
  \frac{\mc F_+(\theta)}{r-\rplus}
  \frac{
    \ell^2\cot^2\theta\,
    (a^4\rplus^2+4M^2q^4)
    (\rplus-\rminus)
    (\rplus^2-q^2)
  }{
    q^2+a^2\cos^2\theta
  }
  \nonumber\\
  &+
  \Ord\!\left[
    \sqrt{
      \frac{\rplus^2-q^2}{r-\rplus}
    }
  \right] \,,
  \qquad
  r\rightarrow\rplus,
  \label{eq:outer-root-main}
\end{align}
where \(\mc F_+(\theta)\)  collects all regular factors in the
angular direction and the subleading order term has the same singular factors (up to a square root). For a general nonextremal rotating configuration, cancellation of the
singular terms selects
\begin{equation}
  q^2=\rplus^2.
  \label{eq:horizon-selected-branch}
\end{equation}
Substituting this condition directly into the complete curvature
invariant confirms that
\begin{equation}
  \left.\Kretsch\right|_{q^2=\rplus^2}
  =\Kretsch_+(\theta)/\sin^2\theta \,,
  \qquad
  0<\theta<\pi \,,
  \label{eq:regular-outer-expansion}
\end{equation}
where \(\Kretsch_+(\theta)\) is finite for every fixed angle.

The same branch is regular at the level of the metric away from
the poles. Indeed, now
\[
\Omega_r
  =
  \sqrt{\frac{r+\rplus}{r-\rminus}} \,,
\]
is finite at \(r=\rplus\), so the bumblebee one-form and the
\(r\theta\) component of the disformal deformation remain finite
there. The remaining Boyer--Lindquist divergence is the familiar
coordinate singularity associated with \(g_{rr}\), and it is
removed in horizon-penetrating coordinates. The surface
\(r=\rplus\) is therefore a Killing horizon on the open angular
interval \(0<\theta<\pi\), generated by
\(\chi=\partial_t+\Omega_H\partial_\phi\), with
\(\Omega_H=r_-/2Ma\).

The statement can be made precise on every domain
\begin{equation}
  \mc D_\epsilon
  =
  \left\{
    r\geq\rplus,\quad
    \epsilon\leq\theta\leq\pi-\epsilon
  \right\},
  \qquad
  \epsilon>0 \,.
  \label{eq:off-axis-domain}
\end{equation}
On the branch \eqref{eq:horizon-selected-branch}, the metric and
bumblebee field are smooth on \(\mc D_\epsilon\) in a
horizon-penetrating chart. No additional curvature singularity is
encountered in the exterior, and at a fixed nonaxial angle the
Kretschmann scalar decays as
\begin{equation}
  \Kretsch
  =
  \frac{4\ell^2}{(1+\ell)^2r^4}
  +
  \Ord(r^{-5}) \,,
  \qquad
  r\rightarrow\infty \,.
  \label{eq:asymptotic-Kretschmann}
\end{equation}

This establishes regularity of the nonpolar exterior and of the
open part of the outer horizon. However, it does not establish regularity
at the two poles, as shown in Eq. (\ref{eq:regular-outer-expansion}). The limits \(\theta\rightarrow0,\pi\) must be analyzed separately, and the
axial geometry will be considered below.

\paragraph{Axial smoothness.---}
Near a regular rotation axis, a smooth axisymmetric one-form must
satisfy
\(\Omega_\theta=\Ord(\sin\theta)\) and
\(\Omega_\phi=\Ord(\sin^2\theta)\), see \ref{app:axis-smoothness}.

For Eq.~\eqref{eq:Kerr-oneform}, however,
\begin{equation}
  \Omega_\theta^2\big|_{\theta=0,\pi}
  =
  a^2+q^2 \,.
  \label{eq:polar-oneform}
\end{equation}
Every rotating member with a real \(q\) therefore has a bumblebee
one-form that fails to extend smoothly to either axis. In the
static sector, the same problem remains whenever \(q\neq0\).

This directly realizes the one-form obstruction, but it does not
yet prove that the metric is curvature singular. Since the
disformal geometry has a nonzero \(r\theta\) component, its
transverse proper distance is controlled by the Schur complement
of the \((r,\theta)\) block. The resulting conicity reads
\begin{equation}
  \alpha^2(r)
  =
  1-
  \frac{
    \ell(a^2+q^2)
  }{
    (1+\ell)(r^2+a^2)
  } \,.
  \label{eq:Kerr-conicity}
\end{equation}
For details, see \ref{app:cartan-varying-conicity}.

Let \(z\) denote proper distance along the exterior axis. For
\(r>\rplus\), its longitudinal derivative is generically nonzero
whenever \(\ell(a^2+q^2)\neq0\). The varying-conicity criterion
then proves that every such open exterior axial segment contains a
frame-curvature singularity.

A direct evaluation of the complete Kretschmann scalar confirms
the corresponding scalar divergence,
\begin{equation}
  \Kretsch
  =
  \frac{C_{\rm ax}(r)}{\sin^2\theta}
  +
  \Ord(\sin^{-1}\theta) \,,
  \qquad
  \theta\rightarrow0,\pi \,,
  \label{eq:axial-Kretschmann}
\end{equation}
where \(C_{\rm ax}(r)\) is generically nonzero  $r\geq r_+$, see \ref{app:complete-axial-Kretschmann} for a complete derivation. The singular axes cannot be regularized by assigning a different
constant period to \(\phi\), because the required period would
vary with position. The defect is therefore not a uniform
cosmic-string-type conical singularity in the C-metric case \cite{Kinnersley:1970zw,Plebanski:1976gy}. 

The varying-conicity obstruction also extends beyond the separated
family. For envelopes generated by complete Hamilton--Jacobi
integrals that are real throughout the exterior, the separation
parameter satisfies
\(0\leq q^2(r,\theta)\leq r_+^2\). Constant conicity compatible
with the nonpolar horizon would instead require
\(q^2(r,0)=r^2\), which violates this bound for every
\(r>r_+\). A globally real horizon-compatible envelope therefore
cannot make the conicity constant along the complete exterior axis,
and the varying-conicity curvature obstruction persists; see
Supplemental Material~\ref{app:separable-obstruction}.

\section{Horizon branches}
\label{sec:horizon-branches}

Axis regularity and horizon regularity impose distinct conditions.
Hence, the parameter \(q\) may remove a nonaxial divergence at the
outer Kerr-like null hypersurface without restoring axial
regularity.

\paragraph{Static sector.---}
For \(a=0\), axial smoothness requires \(q=0\). Independently,
regularity of the inherited Schwarzschild surface \(r=2M\) away
from the axis selects \(q=0\) or \(q^2=4M^2\).

The first branch reproduces the axis-regular
Schwarzschild-like bumblebee exterior
\cite{CasanaEtAl2018}. The second regularizes the nonaxial null
hypersurface but retains the nonsmooth polar component
\(q\,\dd\theta\). Its conicity varies as
\[
\alpha^2(r)
=
1-\frac{\ell q^2}{(1+\ell)r^2}\,,
\]
so it still possesses a curvature-singular axis. Notably, the
static sector already shows that regularity of a null hypersurface
does not imply regularity of the complete metric--bumblebee
configuration.

\paragraph{Nonextremal rotating sector.---}
In the nonextremal rotating sector \(0<|a|<M\), a direct curvature
expansion near the simple outer root shows that it is generically
singular away from the axis. Cancellation of the leading nonaxial
divergence selects Eq.~\eqref{eq:horizon-selected-branch}.
On this branch, the open nonpolar portion of \(r=\rplus\) is a
regular Killing horizon for \(0<\theta<\pi\). Its generator has the
angular velocity
\(\Omega_H=r_-/2Ma\), and the local Hawking temperature is
\[
T_H
=
\frac{\rplus-\rminus}
{8\pi M r_+}\,.
\]
These are local horizon quantities defined away from the singular
poles and do not imply global horizon regularity.

However, regularity of the inner Kerr surface is not preserved.
Direct evaluation gives
\(\Kretsch\sim C_-(\theta)/(r-\rminus)\), with a generically
nonzero coefficient at fixed nonaxial angle, so \(r=\rminus\) is a
null curvature singularity rather than a regular Cauchy horizon.
Therefore, the radial causal structure of the solution differs
substantially from that of Kerr, as discussed in Supplemental
Material~\ref{app:causal-structure}.

Notice that the condition in
Eq.~\eqref{eq:horizon-selected-branch} is a nonaxial
horizon-regularity condition rather than a global regularity
condition. At either pole,
\(\Omega_\theta^2=a^2+\rplus^2>0\), and hence the exterior axes
remain curvature singular.

\paragraph{Extremal sector.---}
For \(|a|=M\), the two Kerr roots coincide and
\(\Delta=(r-M)^2\). In this case, the nonaxial scalar-curvature
divergence at \(r=r_+\) can be absent for a generic real \(q\) in
the allowed range. However, the axial obstruction remains
immediate:
\(\Omega_\theta^2|_{\theta=0,\pi}=M^2+q^2>0\) for every real
\(q\). No real separated extremal member has a smooth rotation
axis. The radial square root is real in an exterior neighborhood
of the degenerate surface only when \(q^2\leq M^2\).

\section{Discussion}
\label{sec:discussion}

We have identified two complementary obstructions to globally
regular rotating constant-norm vector vacua. At an axial fixed
point of a regular nondegenerate bifurcation surface, horizon-boost
and axial invariance force every smooth symmetry-inheriting
one-form to vanish, in conflict with a strictly nonzero constant
norm. Independently, conicity varying along a rotation axis
produces
\(R_{\hat\rho\hat\phi\hat z\hat\phi}\sim\rho^{-1}\)
and precludes a \(C^2\) axial extension. The first obstruction
concerns smoothness of the Lorentz-breaking field, whereas the
second is a geometric curvature obstruction.

The exact Kerr--disformal family realizes both mechanisms. Its
bumblebee one-form violates the local axis-smoothness condition,
and its conicity varies along both open exterior axes. Direct
evaluation further gives
\(\mathcal K\sim\rho^{-2}\), establishing that these axes are
scalar-polynomial curvature singularities. On the distinguished
branch \(q^2=\rplus^2\), the open nonpolar portion of the outer
Killing horizon is regular, whereas the inherited inner root is a
null curvature singularity. This branch therefore defines an exact
nonpolar rotating exterior solution, but not a globally regular
isolated black hole of the complete metric--bumblebee system.

The distinction between metric and full-field regularity is already
present in the static sector. Known Schwarzschild-like bumblebee
configurations remain valid exact exterior solutions with regular
metric horizons, although their standard radial constant-norm
one-forms do not admit smooth extensions through a nondegenerate
bifurcation surface
\cite{CasanaEtAl2018,Chen:2025ypx}. They should therefore be
distinguished from globally smooth bifurcate solutions without
being dismissed as invalid exterior geometries. To the best of our
knowledge, the currently available rotating constant-norm
bumblebee constructions
\cite{DingEtAl2020,PoulisSoares2022,Ovcharenko:2026rvj},
including the related metric-affine solution of
Ref.~\cite{AraujoFilho:2024ykw}, likewise do not satisfy this
stronger global regularity requirement.

Our fixed point result implies a no-go theorem, excluding globally regular
rotating constant-norm bumblebee vacua under symmetry inheritance and the presence of a regular nondegenerate bifurcate Killing horizon. Extremal or one-sided future horizons,
symmetry-noninheriting fields, varying-norm phases, geometries
without axial fixed points, and a dynamical resolution of the polar
region evade at least one assumption of the analysis. Allowing a
nonclosed field does not evade the fixed-point obstruction, but may
open solution sectors beyond the closed disformal construction
considered here. Any globally regular rotating solution, if one
exists, must therefore evade at least one assumption underlying the
two obstructions.

\begin{acknowledgments}
This work was supported by the National Natural Science Foundation
of China under Grants No.~12575048 and No.~11873025.
M.G. was also supported by the BNU Tang Scholar program.

\end{acknowledgments}

\bibliography{references}

\clearpage

\onecolumngrid

\begin{center}
\textbf{\large Supplementary Materials}
\end{center}
\setcounter{equation}{0}
\setcounter{figure}{0}
\setcounter{table}{0}

\setcounter{section}{0}

\makeatletter
\renewcommand{\thesection}{S-\Roman{section}}
\renewcommand{\theequation}{S\arabic{equation}}
\renewcommand{\thefigure}{S\arabic{figure}}
\renewcommand{\thetable}{S\arabic{table}}

\section{Fixed-point proof in an adapted frame}
\label{app:fixed-point-proof}

At \(p\in\mc B\cap\mc A\), choose an orthonormal frame
\(\{e_{\hat0},e_{\hat1},e_{\hat2},e_{\hat3}\}\), with dual
coframe \(\{e^{\hat0},e^{\hat1},e^{\hat2},e^{\hat3}\}\).
The first two vectors span the normal plane to \(\mc B\), while
the last two span \(T_p\mc B\). For a nondegenerate bifurcate
Killing horizon,
\begin{equation}
\nabla_a\chi_b\big|_p
=
\kappa_H
\left(
e^{\hat0}_ae^{\hat1}_b
-
e^{\hat1}_ae^{\hat0}_b
\right),
\qquad
\kappa_H\neq0.
\label{eq:boost-generator}
\end{equation}
Since \(\chi|_p=0\), symmetry inheritance gives
\begin{equation}
0
=
(\Lie_\chi\Omega)_a\big|_p
=
\Omega_b\nabla_a\chi^b\big|_p.
\end{equation}
The boost generator is invertible on the normal plane, and hence
\begin{equation}
\Omega_{\hat0}(p)
=
\Omega_{\hat1}(p)
=
0.
\end{equation}

At the same point, the linearized axial action is
\begin{equation}
\nabla_am_b\big|_p
=
\omega
\left(
e^{\hat2}_ae^{\hat3}_b
-
e^{\hat3}_ae^{\hat2}_b
\right),
\qquad
\omega\neq0.
\label{eq:rotation-generator}
\end{equation}
Because \(m|_p=0\), axial invariance similarly gives
\begin{equation}
0
=
(\Lie_m\Omega)_a\big|_p
=
\Omega_b\nabla_am^b\big|_p,
\end{equation}
and the rotation generator is invertible on \(T_p\mc B\).
Therefore,
\begin{equation}
\Omega_{\hat2}(p)
=
\Omega_{\hat3}(p)
=
0.
\end{equation}
Combining the normal and tangential components yields
\begin{equation}
\Omega|_p=0.
\end{equation}

The proof uses only smoothness, symmetry inheritance, and the two
fixed-point isotropy actions. It requires neither closedness nor a
potential for \(\Omega\), and is independent of the field
equations.

\section{Smooth axisymmetric one-forms}
\label{app:axis-smoothness}

Near a regular axis, introduce smooth Cartesian coordinates
\((x,y)\) on the transverse plane and coordinates \(y^A\) tangent
to the axial fixed-point set, with
\begin{equation}
\rho^2=x^2+y^2.
\end{equation}
The axial Killing field is locally
\begin{equation}
m=x\partial_y-y\partial_x.
\end{equation}
Every smooth invariant one-form can be written as
\begin{align}
\Omega
={}&
\Omega_A(\rho^2,y^B)\dd y^A
+
F(\rho^2,y^B)
(x\,\dd x+y\,\dd y)
\nonumber\\
&+
G(\rho^2,y^B)
(x\,\dd y-y\,\dd x),
\label{eq:smooth-axisymmetric-oneform}
\end{align}
where all coefficient functions are smooth. Using
\begin{equation}
x\,\dd x+y\,\dd y
=
\rho\,\dd\rho,
\qquad
x\,\dd y-y\,\dd x
=
\rho^2\dd\phi,
\end{equation}
one obtains the necessary near-axis behavior
\begin{equation}
\Omega_\rho=\Ord(\rho),
\qquad
\Omega_\phi=\Ord(\rho^2).
\label{eq:smooth-oneform-polar}
\end{equation}

In Boyer--Lindquist-type coordinates adapted to a regular rotation
axis, \(\rho\propto\sin\theta\). Consequently, smoothness at either
axis requires
\begin{equation}
\Omega_\theta=\Ord(\sin\theta),
\qquad
\Omega_\phi=\Ord(\sin^2\theta).
\label{eq:smooth-oneform-BL}
\end{equation}

\section{Cartan-frame proof of the varying-conicity criterion}
\label{app:cartan-varying-conicity}

We give a Cartan-frame derivation of the universal curvature
divergence generated by longitudinally varying conicity. Let
\(m=\partial_\phi\) be the axial Killing field, with
\(\phi\sim\phi+2\pi\), and consider a punctured neighborhood
\(\rho>0\) of a candidate rotation axis. The coordinate \(\rho\)
is proper distance from the axis, while \(z\) is chosen as proper
distance along the axis at the point under consideration.

Let
\begin{equation}
f=\sqrt{g_{\phi\phi}}
\label{eq:cartan-orbit-radius}
\end{equation}
be the proper radius of an azimuthal orbit. We assume that
\begin{equation}
f(\rho,z)
=
\alpha(z)\rho+\rho^2\beta(\rho,z),
\label{eq:cartan-f-expansion}
\end{equation}
where \(\alpha(z)>0\), while \(\beta\) and the derivatives required
below remain bounded as \(\rho\to0\).

We further assume that the quotient geometry obtained by removing
the azimuthal orbits has a regular \(C^2\) limit. The angular mixed
components obey
\begin{equation}
g_{i\phi}
=
\mathcal O(\rho^2),
\qquad
i\in\{\rho,t,z\},
\label{eq:cartan-angular-mixing}
\end{equation}
and the rotational one-form
\begin{equation}
A_i
=
\frac{g_{i\phi}}{g_{\phi\phi}}
\label{eq:cartan-A-definition}
\end{equation}
is assumed to have bounded coefficients and bounded derivatives
through the order required by the curvature calculation. These
conditions allow regular stationary frame dragging while excluding
an independent singularity hidden in the angular mixed terms.

\subsection{Azimuthal decomposition}
\label{app:cartan-decomposition}

Away from the axis, the metric admits the exact local decomposition
\begin{equation}
\dd s^2
=
q_{ij}(y)\dd y^i\dd y^j
+
f^2
\left(
\dd\phi+A_i\dd y^i
\right)^2,
\label{eq:cartan-metric-decomposition}
\end{equation}
where
\begin{equation}
y^i=(\rho,t,z),
\end{equation}
and
\begin{equation}
q_{ij}
=
g_{ij}
-
\frac{g_{i\phi}g_{j\phi}}{g_{\phi\phi}}.
\label{eq:cartan-quotient-metric}
\end{equation}
The tensor \(q_{ij}\) is the Lorentzian metric on the local quotient
by the axial \(U(1)\) action.

Because \(\rho\) is a Gaussian normal coordinate for the quotient
geometry, \(q\) can locally be written as
\begin{equation}
q
=
\dd\rho^2
+
h_{AB}(\rho,x)\dd x^A\dd x^B,
\qquad
x^A=(t,z).
\label{eq:cartan-Gaussian-quotient}
\end{equation}
Thus
\begin{equation}
q_{\rho\rho}=1,
\qquad
q_{\rho A}=0.
\label{eq:cartan-Gaussian-conditions}
\end{equation}
We assume that \(h_{AB}\), its inverse, and their first two
derivatives remain bounded as \(\rho\to0\).

Choose an orthonormal coframe
\(\{e^{\hat i}\}\) for the quotient metric, where
\(\hat i\in\{\hat0,\hat\rho,\hat z\}\), with
\begin{equation}
e^{\hat\rho}=\dd\rho.
\label{eq:cartan-radial-coframe}
\end{equation}
At the axial point under consideration, \(e_{\hat z}\) is chosen
to be the unit spacelike vector tangent to the axis. The azimuthal
coframe element is
\begin{equation}
e^{\hat\phi}
=
f\left(\dd\phi+A\right),
\qquad
A=A_i\dd y^i.
\label{eq:cartan-angular-coframe}
\end{equation}
The full metric is then
\begin{equation}
\dd s^2
=
\eta_{\hat i\hat j}
e^{\hat i}e^{\hat j}
+
\left(e^{\hat\phi}\right)^2.
\label{eq:cartan-orthonormal-metric}
\end{equation}

\subsection{Connection one-forms}
\label{app:cartan-angular-connection}

Taking the exterior derivative of
Eq.~\eqref{eq:cartan-angular-coframe} gives
\begin{align}
\dd e^{\hat\phi}
&=
\dd f\wedge(\dd\phi+A)
+
f\,\dd A
\nonumber\\
&=
\dd\ln f\wedge e^{\hat\phi}
+
fF,
\label{eq:cartan-dephi}
\end{align}
where \(F=\dd A\). In the quotient orthonormal frame, write
\begin{equation}
\dd\ln f
=
u_{\hat i}e^{\hat i},
\qquad
u_{\hat i}
=
e_{\hat i}(\ln f),
\label{eq:cartan-u-definition}
\end{equation}
and
\begin{equation}
F
=
\frac12F_{\hat i\hat j}
e^{\hat i}\wedge e^{\hat j}.
\label{eq:cartan-F-expansion}
\end{equation}
It follows that
\begin{equation}
\dd e^{\hat\phi}
=
u_{\hat i}
e^{\hat i}\wedge e^{\hat\phi}
+
\frac{f}{2}
F_{\hat i\hat j}
e^{\hat i}\wedge e^{\hat j}.
\label{eq:cartan-dephi-frame}
\end{equation}

Let \(\overline{\omega}^{\hat i}{}_{\hat j}\) denote the
Levi-Civita connection one-forms of the quotient metric \(q\).
The torsion-free first Cartan equation,
\begin{equation}
\dd e^{\hat a}
+
\omega^{\hat a}{}_{\hat b}
\wedge e^{\hat b}
=
0,
\label{eq:first-Cartan-equation}
\end{equation}
is solved by
\begin{align}
\omega^{\hat\phi}{}_{\hat i}
&=
u_{\hat i}e^{\hat\phi}
+
\frac{f}{2}
F_{\hat i\hat j}e^{\hat j},
\label{eq:cartan-angular-connection}
\\
\omega^{\hat i}{}_{\hat j}
&=
\overline{\omega}^{\hat i}{}_{\hat j}
-
\frac{f}{2}
F^{\hat i}{}_{\hat j}e^{\hat\phi}.
\label{eq:cartan-base-connection}
\end{align}
The remaining connection components follow from
\(\omega_{\hat a\hat b}=-\omega_{\hat b\hat a}\).

Since \(F_{\hat i\hat j}\) and its required derivatives are
bounded, while \(f=\mathcal O(\rho)\), every explicit
twist-dependent correction in
Eqs.~\eqref{eq:cartan-angular-connection} and
\eqref{eq:cartan-base-connection} is \(\mathcal O(\rho)\).
A radial derivative of such a term is at most
\(\mathcal O(1)\). Its product with the standard polar connection
\begin{equation}
e_{\hat\rho}(\ln f)
=
\frac{1}{\rho}
+
\mathcal O(1)
\end{equation}
is also at most \(\mathcal O(1)\). Consequently, the rotational
field \(F\) can modify the bounded part of the curvature considered
below, but it cannot generate or cancel its universal
\(1/\rho\) term.

\subsection{Curvature involving the shrinking orbit}
\label{app:cartan-angular-curvature}

The curvature two-forms are defined by the second Cartan equation,
\begin{equation}
\mathcal R^{\hat a}{}_{\hat b}
=
\dd\omega^{\hat a}{}_{\hat b}
+
\omega^{\hat a}{}_{\hat c}
\wedge
\omega^{\hat c}{}_{\hat b},
\label{eq:second-Cartan-equation}
\end{equation}
with
\begin{equation}
\mathcal R^{\hat a}{}_{\hat b}
=
\frac12
R^{\hat a}{}_{\hat b\hat c\hat d}
e^{\hat c}\wedge e^{\hat d}.
\label{eq:curvature-two-form}
\end{equation}

Substituting the connection one-forms above into
Eq.~\eqref{eq:second-Cartan-equation}, the curvature components
with two azimuthal indices can be organized as
\begin{equation}
R_{\hat i\hat\phi\hat j\hat\phi}
=
-
\frac{\mathcal H_{\hat i\hat j}}{f}
+
\mathcal T_{\hat i\hat j},
\label{eq:cartan-fiber-curvature}
\end{equation}
where
\begin{equation}
\mathcal H_{ij}
=
\overline{\nabla}_i\overline{\nabla}_j f
\end{equation}
is the Hessian of \(f\) with respect to the quotient metric, and
\(\overline{\nabla}\) denotes the Levi-Civita connection of \(q\).
The tensor \(\mathcal T_{\hat i\hat j}\) contains the terms
involving \(F\). The estimates above imply
\begin{equation}
\mathcal T_{\hat i\hat j}
=
\mathcal O(1)
\qquad
(\rho\to0).
\label{eq:cartan-twist-bounded}
\end{equation}

For completeness, the Hessian structure in
Eq.~\eqref{eq:cartan-fiber-curvature} can be seen directly by
retaining only the potentially singular part of the connection:
\begin{equation}
\omega^{\hat\phi}{}_{\hat j}
=
u_{\hat j}e^{\hat\phi}
+
\mathcal O(\rho).
\label{eq:cartan-leading-connection}
\end{equation}
The coefficient of
\(e^{\hat i}\wedge e^{\hat\phi}\) in
\(\mathcal R^{\hat\phi}{}_{\hat j}\) is then
\begin{equation}
-
\left(
\overline{\nabla}_{\hat i}u_{\hat j}
+
u_{\hat i}u_{\hat j}
\right)
+
\mathcal O(1).
\label{eq:cartan-u-curvature}
\end{equation}
Since \(u_{\hat j}=f^{-1}e_{\hat j}f\),
\begin{equation}
\overline{\nabla}_{\hat i}u_{\hat j}
+
u_{\hat i}u_{\hat j}
=
\frac{
\overline{\nabla}_{\hat i}
\overline{\nabla}_{\hat j}f
}{f},
\label{eq:cartan-Hessian-identity}
\end{equation}
which gives Eq.~\eqref{eq:cartan-fiber-curvature}. The overall sign
depends on the convention chosen for the Riemann tensor.

\subsection{Near-axis behavior}
\label{app:cartan-near-axis-evaluation}

We now evaluate the mixed component with
\(\hat i=\hat\rho\) and \(\hat j=\hat z\). Since \(\rho\) is a
Gaussian normal coordinate,
\begin{equation}
\overline{\nabla}_{e_{\hat\rho}}e_{\hat\rho}=0.
\label{eq:cartan-radial-geodesic}
\end{equation}
Choose \(e_{\hat z}\) on the axis and parallel transport it along
the radial Gaussian geodesics:
\begin{equation}
\overline{\nabla}_{e_{\hat\rho}}e_{\hat z}=0.
\label{eq:cartan-Fermi-transport}
\end{equation}
This is a regular choice of orthonormal frame and places no
additional restriction on the quotient geometry.

For a scalar \(f\),
\begin{equation}
\overline{\nabla}_{\hat\rho}
\overline{\nabla}_{\hat z}f
=
e_{\hat\rho}\!\left(e_{\hat z}f\right)
-
\left(
\overline{\nabla}_{e_{\hat\rho}}e_{\hat z}
\right)(f).
\label{eq:cartan-Hessian-definition}
\end{equation}
Equation~\eqref{eq:cartan-Fermi-transport} therefore gives
\begin{equation}
\overline{\nabla}_{\hat\rho}
\overline{\nabla}_{\hat z}f
=
e_{\hat\rho}\!\left(e_{\hat z}f\right).
\label{eq:cartan-Hessian-Fermi}
\end{equation}

Using Eq.~\eqref{eq:cartan-f-expansion} and the fact that
\(e_{\hat z}=\partial_z\) on the axis,
\begin{equation}
e_{\hat z}f
=
\partial_z\alpha\,\rho
+
\mathcal O(\rho^2).
\label{eq:cartan-z-derivative-f}
\end{equation}
Hence
\begin{equation}
\overline{\nabla}_{\hat\rho}
\overline{\nabla}_{\hat z}f
=
\partial_z\alpha
+
\mathcal O(\rho).
\label{eq:cartan-mixed-Hessian}
\end{equation}
Meanwhile,
\begin{equation}
f
=
\alpha\rho
\left[
1+\mathcal O(\rho)
\right].
\label{eq:cartan-f-factorized}
\end{equation}
Therefore,
\begin{equation}
\frac{
\overline{\nabla}_{\hat\rho}
\overline{\nabla}_{\hat z}f
}{f}
=
\frac{\partial_z\ln\alpha}{\rho}
+
\mathcal O(1).
\label{eq:cartan-Hessian-ratio}
\end{equation}

Combining Eqs.~\eqref{eq:cartan-fiber-curvature},
\eqref{eq:cartan-twist-bounded}, and
\eqref{eq:cartan-Hessian-ratio}, we obtain
\begin{equation}
R_{\hat\rho\hat\phi\hat z\hat\phi}
=
-\frac{\partial_z\ln\alpha}{\rho}
+
\mathcal O(1).
\label{eq:cartan-final-result}
\end{equation}
Equation~\eqref{eq:cartan-final-result} isolates the universal
\(1/\rho\) curvature term controlled solely by the longitudinal
variation of the conicity. Other Hessian components may also
contribute at order \(1/\rho\), depending on the subleading
orbit-radius and quotient data, but they cannot cancel the mixed
component in Eq.~\eqref{eq:cartan-final-result}.

\subsection{Consequence for axial regularity}
\label{app:cartan-regularity-consequence}

At an axial point \(z=z_0\) satisfying
\(\partial_z\alpha(z_0)\neq0\),
Eq.~\eqref{eq:cartan-final-result} implies
\begin{equation}
\left|
R_{\hat\rho\hat\phi\hat z\hat\phi}
\right|
\sim
\frac{
|\partial_z\ln\alpha(z_0)|
}{\rho}
\longrightarrow\infty
\qquad
(\rho\to0).
\label{eq:cartan-curvature-divergence}
\end{equation}
This divergence is measured in a unit orthonormal frame. It
persists in every orthonormal frame related to the adapted frame by
a Lorentz transformation that remains bounded as the axis is
approached, and therefore cannot be removed by a regular coordinate
or frame transformation.

A \(C^2\) extension through the axis would have continuous and
finite Riemann components in a bounded orthonormal frame.
Consequently,
\begin{equation}
\partial_z\alpha(z_0)\neq0
\quad\Longrightarrow\quad
\text{no \(C^2\) extension through the axis at \(z_0\)}.
\label{eq:cartan-no-C2-extension}
\end{equation}

The component in
Eq.~\eqref{eq:cartan-curvature-divergence} lies entirely in the
spatial
\((\hat\rho,\hat z,\hat\phi)\) sector. Its Riemann-symmetry
multiplicity gives the algebraic contribution
\begin{equation}
\mathcal K_{\rm cone}
=
\frac{8}{\rho^2}
\left(
\partial_z\ln\alpha
\right)^2
\label{eq:cartan-universal-K-contribution}
\end{equation}
to the leading near-axis Kretschmann contraction. This expression
isolates the contribution of the universal mixed component; it need
not equal the complete leading coefficient in a general stationary
geometry.

For example, in the local untwisted warped-product model
\begin{equation}
\dd s^2_{\rm loc}
=
-\dd t^2
+
\dd\rho^2
+
\dd z^2
+
\alpha^2(z)\rho^2\dd\phi^2,
\end{equation}
the complete Kretschmann scalar is
\begin{equation}
\mathcal K_{\rm loc}
=
\frac{8}{\rho^2}
\left(
\partial_z\ln\alpha
\right)^2
+
4
\left(
\frac{\partial_z^2\alpha}{\alpha}
\right)^2.
\label{eq:local-conicity-Kretschmann}
\end{equation}

In a general stationary-axisymmetric geometry, other
\(1/\rho\) Riemann components may contribute to the scalar
contraction at the same order. When the complete invariant admits
an expansion of the form
\begin{equation}
\mathcal K
=
\frac{C_{\rm full}(z)}{\rho^2}
+
\mathcal O(\rho^{-1}),
\label{eq:general-conicity-Kretschmann}
\end{equation}
its coefficient may be decomposed as
\begin{equation}
C_{\rm full}(z)
=
8
\left(
\partial_z\ln\alpha
\right)^2
+
C_{\rm rem}(z),
\label{eq:general-conicity-coefficient}
\end{equation}
where \(C_{\rm rem}\) collects the remaining same-order
contributions. The Cartan analysis fixes the universal
varying-conicity term, while the complete scalar coefficient must
be evaluated for the geometry under consideration.

The converse is not asserted. If
\(\partial_z\alpha=0\) at an isolated axial point, the universal
mixed-curvature term vanishes there, but other curvature components
or a nonuniform joint limit may still obstruct regularity.

\subsection{Applicability to the Kerr--disformal axis}
\label{app:applicability-cartan}

We verify that the Kerr--disformal metric satisfies the hypotheses
of the Cartan-frame varying-conicity criterion on every open
exterior axial segment. Let
\begin{equation}
R=r^2+a^2,
\qquad
P=r^2-q^2,
\qquad
Q_0=a^2+q^2,
\end{equation}
and define
\begin{equation}
D_0=R+\ell P.
\end{equation}
At \(\theta=0\), the \((r,\theta)\) block has
\begin{align}
g_{rr}
&=
\frac{D_0}{\Delta}
+
\mathcal O(\theta^2),
\\
g_{r\theta}
&=
\frac{\ell\sqrt{P Q_0}}{\sqrt{\Delta}}
+
\mathcal O(\theta^2),
\\
g_{\theta\theta}
&=
R+\ell Q_0+\mathcal O(\theta^2).
\end{align}
For \(r>r_+\), \(P>0\), \(1+\ell>0\), and \(D_0>0\), these
coefficients and their first two derivatives remain bounded on
every compact open exterior axial segment.

The proper transverse coefficient is the Schur complement
\begin{equation}
H
=
g_{\theta\theta}
-
\frac{g_{r\theta}^2}{g_{rr}}
=
\frac{
(1+\ell)\Sigma^2
}{
\Sigma+\ell(r^2-q^2)
},
\end{equation}
whose axial value is
\begin{equation}
H_0
=
\frac{(1+\ell)R^2}{D_0}.
\end{equation}
Gaussian proper distance from the axis therefore satisfies
\begin{equation}
\theta
=
\frac{\rho}{\sqrt{H_0}}
+
\mathcal O(\rho^2),
\qquad
r=r_0+\mathcal O(\rho).
\label{eq:Gaussian-axis-relation}
\end{equation}
The linear displacement of \(r\) reflects the nonzero
\(g_{r\theta}\) component and allows both the quotient data and
\(f\) to contain subleading terms odd in \(\rho\).

The azimuthal orbit radius is
\begin{equation}
f
=
\sqrt{g_{\phi\phi}}
=
\sqrt R\,\theta
+
\mathcal O(\theta^3).
\end{equation}
Using Eq.~\eqref{eq:Gaussian-axis-relation}, one obtains
\begin{equation}
f(\rho,z)
=
\alpha(z)\rho
+
\mathcal O(\rho^2),
\end{equation}
where
\begin{equation}
\alpha^2(r)
=
\frac{R}{H_0}
=
1-
\frac{
\ell(a^2+q^2)
}{
(1+\ell)(r^2+a^2)
}.
\label{eq:Kerr-axis-conicity}
\end{equation}

The stationary angular cross term obeys
\begin{equation}
g_{t\phi}
=
-\frac{2Mar}{R}\theta^2
+
\mathcal O(\theta^4)
=
\mathcal O(\rho^2),
\end{equation}
while \(g_{r\phi}=g_{\theta\phi}=0\). Hence the rotational one-form
in the decomposition
\begin{equation}
\dd s^2
=
q_{ij}\dd y^i\dd y^j
+
f^2(\dd\phi+A)^2
\end{equation}
is bounded:
\begin{equation}
A_t
=
-\frac{2Mar}{R^2}
+
\mathcal O(\rho),
\qquad
A_z=A_\rho=0.
\end{equation}
It follows that
\begin{equation}
f\,\dd A
=
\mathcal O(\rho),
\end{equation}
so the stationary twist contributes at most bounded terms to the
mixed orthonormal curvature.

Finally, proper distance along the exterior axis satisfies
\begin{equation}
\frac{\dd z}{\dd r}
=
\sqrt{\frac{D_0}{\Delta}},
\end{equation}
and therefore
\begin{equation}
\partial_z\alpha
=
\sqrt{\frac{\Delta}{D_0}}\,
\frac{
\ell(a^2+q^2)r
}{
(1+\ell)\alpha(r)(r^2+a^2)^2
}.
\label{eq:Kerr-conicity-z-derivative}
\end{equation}
Under the exterior regularity conditions
\begin{equation}
r>r_+,
\qquad
D_0>0,
\qquad
\alpha^2>0,
\end{equation}
this derivative is nonzero whenever
\(\ell(a^2+q^2)\neq0\). The Cartan-frame criterion consequently
gives
\begin{equation}
R_{\hat\rho\hat\phi\hat z\hat\phi}
=
-\frac{\partial_z\ln\alpha}{\rho}
+
\mathcal O(1),
\end{equation}
and excludes a \(C^2\) extension through every such open exterior
axial point.

The conclusion applies to \(r>r_+\). At the horizon poles, the
Boyer--Lindquist chart degenerates and
\(\partial_z\alpha\to0\) on the horizon-selected branch. Their
regularity must instead be tested in a horizon-penetrating chart or
through a direct joint near-horizon--near-axis curvature expansion.

\section{Complete leading axial Kretschmann coefficient}
\label{app:complete-axial-Kretschmann}
Now let us derive the complete coefficient of the
leading axial divergence of the Kretschmann scalar for the
Kerr-disformal metric. More precisely, we calculate
\begin{equation}
  C_{\rm ax}(r)
  =
  \lim_{\theta\to0}
  \sin^2\theta\,
  R_{\mu\nu\rho\sigma}R^{\mu\nu\rho\sigma}\,.
  \label{eq:Cax-definition}
\end{equation}
The same coefficient is obtained at the south axis
\(\theta\to\pi\).

The calculation applies to an open exterior axial segment on
which
\begin{equation}
  \Delta>0\,,
  \qquad
  r^2-q^2>0 \,,
  \qquad
  1+\ell>0 \,.
  \label{eq:axial-calculation-domain}
\end{equation}
The horizon limit of the selected branch
\(q^2=r_+^2\) will be taken only after the exterior coefficient
has been obtained.

\subsection{Quotient geometry and azimuthal radius}
\label{app:K-quotient-geometry}


 Recall that away from the axis,
the metric admits the exact decomposition
\begin{equation}
  \dd s^2
  =
  q_{ij}\dd y^i\dd y^j
  +
  f^2
  \left(
    \dd\phi+A_i\dd y^i
  \right)^2,
  \qquad
  y^i=(t,r,\theta),
  \label{eq:K-quotient-decomposition}
\end{equation}
where $  f=\sqrt{g_{\phi\phi}}$ is the proper radius of an azimuthal orbit and
\begin{equation}
  A_i
  =
  \frac{g_{i\phi}}{g_{\phi\phi}}\,,
  \qquad
  q_{ij}
  =
  g_{ij}
  -
  \frac{g_{i\phi}g_{j\phi}}{g_{\phi\phi}} \,.
  \label{eq:K-quotient-definitions}
\end{equation}

For the Kerr-disformal metric, \(g_{r\phi}=g_{\theta\phi}=0\),
and only \(A_t\) is nonzero. Near the axis,
\(g_{t\phi}=\mathcal O(\sin^2\theta)\) and
\(g_{\phi\phi}=\mathcal O(\sin^2\theta)\), so \(A_t\) and its
required derivatives remain bounded. The rotational field
\(F=\dd A\) therefore contributes only bounded terms to the
curvature components relevant below.

The quotient metric at the north axis is
\begin{equation}
  q_{ij}\dd y^i\dd y^j
  =
  -\frac{\Delta}{R}\dd t^2
  +
  E\,\dd r^2
  +
  2F_0\,\dd r\,\dd\theta
  +
  G\,\dd\theta^2,
  \label{eq:K-axis-quotient}
\end{equation}
where
\begin{equation}
  E=\frac{D_0}{\Delta},
  \qquad
  F_0=\frac{\ell\sqrt{PQ_0}}{\sqrt{\Delta}},
  \qquad
  G=R+\ell Q_0 \,.
  \label{eq:K-axis-block}
\end{equation}
The symbol \(F_0\) denotes the \(r\theta\) metric coefficient and
should not be confused with the rotational two-form \(F=\dd A\).

The determinant of the spatial \((r,\theta)\) block is
\begin{align}
  EG-F_0^2
  &=
  \frac{
    D_0(R+\ell Q_0)-\ell^2PQ_0
  }{\Delta}
  \nonumber\\
  &=
  \frac{(1+\ell)R^2}{\Delta} \,,
  \label{eq:K-spatial-determinant}
\end{align}
where we used \(P+Q_0=R\). The nonzero inverse quotient components
on the axis are therefore
\begin{align}
  q^{tt}
  &=
  -\frac{R}{\Delta}\,,
  \label{eq:K-inverse-tt}
  \\
  q^{rr}
  &=
  \frac{(R+\ell Q_0)\Delta}
  {(1+\ell)R^2}\,,
  \label{eq:K-inverse-rr}
  \\
  q^{r\theta}
  &=
  -\frac{
    \ell\sqrt{PQ_0\Delta}
  }{
    (1+\ell)R^2
  } \,,
  \label{eq:K-inverse-rtheta}
  \\
  q^{\theta\theta}
  &=
  \frac{D_0}{(1+\ell)R^2} \,.
  \label{eq:K-inverse-thetatheta}
\end{align}

Because the disformal deformation does not modify
\(g_{\phi\phi}\), the azimuthal radius is
\begin{equation}
  f
  =
  \sin\theta
  \sqrt{
    \frac{
      R^2-a^2\Delta\sin^2\theta
    }{
      r^2+a^2\cos^2\theta
    }
  } \,.
  \label{eq:K-exact-f}
\end{equation}
Its near-axis expansion is
\begin{equation}
  f
  =
  \sqrt{R}\,\theta+\mathcal O(\theta^3)
  =
  \sqrt{R}\,\sin\theta
  +
  \mathcal O(\sin^3\theta) \,.
  \label{eq:K-f-axis-expansion}
\end{equation}
Consequently, on the north axis,
\begin{equation}
  f=0\,,
  \qquad
  \partial_t f=0 \,,
  \qquad
  \partial_r f=0 \,,
  \qquad
  \partial_\theta f=\sqrt R \,.
  \label{eq:K-first-derivatives-f}
\end{equation}
The only nonzero second coordinate derivative needed at leading
order is
\begin{equation}
  \partial_r\partial_\theta f
  =
  \frac{r}{\sqrt R} \,.
  \label{eq:K-mixed-second-derivative}
\end{equation}

\subsection{Quotient-space Hessian}
\label{app:K-Hessian}

 Since only \(\partial_\theta f=\sqrt R\) is nonzero
on the axis,
\begin{equation}
  \mathcal H_{ij}
  =
  \partial_i\partial_jf
  -
  \overline{\Gamma}^{\theta}{}_{ij}\sqrt R \,.
  \label{eq:K-Hessian-axis}
\end{equation}

We now calculate its independent nonzero components.

\paragraph{Time component.---}
On the axis,
\begin{equation}
  q_{tt}=-\frac{\Delta}{R} \,,
  \qquad
  \partial_rq_{tt}
  =
  -\frac{2M(r^2-a^2)}{R^2} \,.
  \label{eq:K-qtt-derivative}
\end{equation}
Stationarity gives
\begin{equation}
  \overline{\Gamma}^{\theta}{}_{tt}
  =
  -\frac12q^{\theta r}\partial_rq_{tt} \,.
\end{equation}
It follows that
\begin{equation}
  \mathcal H_{tt}
  =
  \frac{
    \ell M(r^2-a^2)\sqrt{PQ_0\Delta}
  }{
    (1+\ell)R^{7/2}
  } \,.
  \label{eq:K-Htt}
\end{equation}

\paragraph{Radial component.---}
Since the first \(\theta\)-derivatives of the quotient metric
coefficients vanish on the axis,
\begin{equation}
  \overline{\Gamma}^{\theta}{}_{rr}
  =
  \frac12q^{\theta r}\partial_rE
  +
  q^{\theta\theta}\partial_rF_0 \,.
  \label{eq:K-Gamma-theta-rr}
\end{equation}
Using Eqs.~\eqref{eq:K-axis-block},
\eqref{eq:K-inverse-rtheta}, and
\eqref{eq:K-inverse-thetatheta}, one obtains
\begin{equation}
  \mathcal H_{rr}
  =
  -\frac{
    \ell rQ_0\sqrt{PQ_0\Delta}
  }{
    (1+\ell)R^{3/2}P\Delta
  } \,.
  \label{eq:K-Hrr}
\end{equation}

\paragraph{Mixed spatial component.---}
The relevant connection coefficient is
\begin{equation}
  \overline{\Gamma}^{\theta}{}_{r\theta}
  =
  \frac12q^{\theta\theta}\partial_rG
  =
  rq^{\theta\theta}.
  \label{eq:K-Gamma-theta-rtheta}
\end{equation}
Using
\(\partial_r\partial_\theta f=r/\sqrt R\), we find
\begin{align}
  \mathcal H_{r\theta}
  &=
  \frac{r}{\sqrt R}
  -
  rq^{\theta\theta}\sqrt R
  \nonumber=
  \frac{r}{\sqrt R}
  \left[
    1-\frac{D_0}{(1+\ell)R}
  \right]
  \nonumber=
  \frac{
    \ell rQ_0
  }{
    (1+\ell)R^{3/2}
  } \,.
  \label{eq:K-Hrtheta}
\end{align}

\paragraph{Angular component.---}
The corresponding connection coefficient is
\begin{equation}
  \overline{\Gamma}^{\theta}{}_{\theta\theta}
  =
  -\frac12q^{\theta r}\partial_rG
  =
  -rq^{\theta r}.
  \label{eq:K-Gamma-theta-thetatheta}
\end{equation}
Therefore,
\begin{equation}
  \mathcal H_{\theta\theta}
  =
  -\frac{
    \ell r\sqrt{PQ_0\Delta}
  }{
    (1+\ell)R^{3/2}
  } \,.
  \label{eq:K-Hthetatheta}
\end{equation}

All time--space Hessian components vanish:
\begin{equation}
  \mathcal H_{tr}
  =
  \mathcal H_{t\theta}
  =
  0 \,.
  \label{eq:K-mixed-time-Hessian}
\end{equation}

\subsection{Hessian contraction}
\label{app:K-Hessian-contraction}

The leading curvature components containing the shrinking
azimuthal direction are
\begin{equation}
  R_{\hat i\hat\phi\hat j\hat\phi}
  =
  -
  \frac{
    \mathcal H_{\hat i\hat j}
  }{f}
  +
  \mathcal O(1) \,.
  \label{eq:K-leading-fiber-curvature}
\end{equation}
The bounded rotational terms do not modify the coefficient of the
\(1/f\) divergence.

The required quotient contraction is
\begin{equation}
  \mathcal H^2
  \equiv
  \mathcal H_{ij}\mathcal H^{ij}
  =
  q^{ik}q^{jl}
  \mathcal H_{ij}\mathcal H_{kl} \,.
  \label{eq:K-Hessian-square}
\end{equation}

The time contribution is
\begin{equation}
  (q^{tt})^2\mathcal H_{tt}^2
  =
  \frac{
    \ell^2M^2(r^2-a^2)^2PQ_0
  }{
    (1+\ell)^2 R^5\Delta
  } \,.
  \label{eq:K-time-contraction}
\end{equation}

For the spatial block, define
\begin{equation}
  Q_{\rm sp}
  =
  \begin{pmatrix}
    q^{rr} & q^{r\theta}\\
    q^{r\theta} & q^{\theta\theta}
  \end{pmatrix},
  \qquad
  \mathcal H_{\rm sp}
  =
  \begin{pmatrix}
    \mathcal H_{rr} & \mathcal H_{r\theta}\\
    \mathcal H_{r\theta} & \mathcal H_{\theta\theta}
  \end{pmatrix}.
  \label{eq:K-spatial-matrices}
\end{equation}
The spatial contraction is
\begin{equation}
  \mathcal H_{{\rm sp}\,ij}
  \mathcal H_{\rm sp}^{\,ij}
  =
  \operatorname{Tr}
  \left[
    \left(
      Q_{\rm sp}\mathcal H_{\rm sp}
    \right)^2
  \right].
  \label{eq:K-spatial-trace}
\end{equation}

Using Eqs.~\eqref{eq:K-inverse-rr}--%
\eqref{eq:K-inverse-thetatheta} and
\eqref{eq:K-Hrr}--\eqref{eq:K-Hthetatheta}, the matrix product
simplifies to
\begin{equation}
  Q_{\rm sp}\mathcal H_{\rm sp}
  =
  \frac{
    \ell r
  }{
    (1+\ell)R^{5/2}
  }
  \begin{pmatrix}
    -\dfrac{Q_0\sqrt{PQ_0\Delta}}{P}
    &
    Q_0\Delta
    \\[2mm]
    Q_0
    &
    -\sqrt{PQ_0\Delta}
  \end{pmatrix}.
  \label{eq:K-QH-matrix}
\end{equation}
Its squared trace is
\begin{equation}
  \operatorname{Tr}
  \left[
    \left(
      Q_{\rm sp}\mathcal H_{\rm sp}
    \right)^2
  \right]
  =
  \frac{
    \ell^2r^2Q_0\Delta
  }{
    (1+\ell)^2R^3P
  } \,.
  \label{eq:K-spatial-contraction-result}
\end{equation}

Combining the time and spatial contributions gives
\begin{equation}
  \mathcal H^2
  =
  \frac{
    \ell^2Q_0
  }{
    (1+\ell)^2
  }
  \left[
    \frac{
      r^2\Delta
    }{
      R^3P
    }
    +
    \frac{
      M^2(r^2-a^2)^2P
    }{
      R^5\Delta
    }
  \right] \,.
  \label{eq:K-complete-Hessian-square}
\end{equation}

\subsection{Leading Kretschmann divergence}
\label{app:K-leading-divergence}

The quotient curvature and all twist-dependent terms are bounded
under the assumptions stated above. The complete
\(1/f^2\) contribution to the Kretschmann scalar therefore comes
from Eq.~\eqref{eq:K-leading-fiber-curvature}. Accounting for the
algebraic symmetries of the Riemann tensor gives
\begin{equation}
  \Kretsch
  =
  \frac{4\mathcal H^2}{f^2}
  +
  \mathcal O(f^{-1}) \,.
  \label{eq:K-leading-general}
\end{equation}
Since
\begin{equation}
  f^2
  =
  R\sin^2\theta
  +
  \mathcal O(\sin^4\theta) \,,
  \label{eq:K-f-square-axis}
\end{equation}
we obtain
\begin{equation}
  \Kretsch
  =
  \frac{
    C_{\rm ax}(r)
  }{
    \sin^2\theta
  }
  +
  \mathcal O(\sin^{-1}\theta) \,,
  \qquad
  \theta\to0,\pi \,,
  \label{eq:K-complete-axis-expansion}
\end{equation}
where
\begin{align}
  C_{\rm ax}(r)
  ={}&
  \frac{
    4\ell^2(a^2+q^2)
  }{
    (1+\ell)^2
  }
  \Bigg[
    \frac{
      r^2\Delta
    }{
      (r^2+a^2)^4(r^2-q^2)
    }
    +
    \frac{
      M^2(r^2-a^2)^2(r^2-q^2)
    }{
      (r^2+a^2)^6\Delta
    }
  \Bigg] \,.
  \label{eq:K-complete-Cax}
\end{align}
This gives the complete coefficient of
the leading \(1/\sin^2\theta\) divergence. In the exterior domain \eqref{eq:axial-calculation-domain}, both
terms in Eq.~\eqref{eq:K-complete-Cax} are nonnegative and hence
\begin{equation}
  C_{\rm ax}(r)>0
  \label{eq:K-Cax-positive}
\end{equation}
whenever
\begin{equation}
  \ell\neq0,
  \qquad
  a^2+q^2>0.
\end{equation}
The open exterior axes are therefore scalar-polynomial curvature
singularities.

Several limits provide useful checks:
\begin{equation}
  C_{\rm ax}\big|_{\ell=0}=0,
  \qquad
  C_{\rm ax}\big|_{a=q=0}=0.
  \label{eq:K-Cax-checks}
\end{equation}
The first recovers the regular Kerr axis, while the second
recovers the regular static \(q=0\) branch.

\subsection{Horizon-selected branch}
\label{app:K-horizon-pole}

For the nonextremal horizon-selected branch
\begin{equation}
  q^2=r_+^2 \,,
  \qquad
  \Delta=(r-r_+)(r-r_-) \,,
  \label{eq:K-selected-branch}
\end{equation}
the two apparently singular ratios in
Eq.~\eqref{eq:K-complete-Cax} have finite outer-horizon limits:
\begin{equation}
  \frac{\Delta}{r^2-r_+^2}
  \longrightarrow
  \frac{r_+-r_-}{2r_+} \,,
  \qquad
  \frac{r^2-r_+^2}{\Delta}
  \longrightarrow
  \frac{2r_+}{r_+-r_-} \,.
  \label{eq:K-horizon-ratios}
\end{equation}
It follows that
\begin{equation}
  C_H
  \equiv
  \lim_{r\to r_+}C_{\rm ax}(r)
  =
  \frac{
    4\ell^2r_+(r_+-r_-)
  }{
    (1+\ell)^2
    (r_+^2+a^2)^3
  } \,.
  \label{eq:K-horizon-pole-coefficient}
\end{equation}
We used
\(a^2=r_+r_-\) and
\(r_+^2+a^2=2Mr_+\) to obtain the final form.

For a nonextremal black hole, \(r_+>r_-\), and hence
\(C_H>0\) whenever \(\ell\neq0\). The outer horizon is therefore
regular only at fixed nonaxial angles. Its north and south poles
remain curvature singularities, with
\begin{equation}
  \Kretsch
  \sim
  \frac{C_H}{\sin^2\theta} \,,
  \qquad
  \theta\to0,\pi \,.
  \label{eq:K-horizon-pole-divergence}
\end{equation}

The axial and horizon limits are consequently nonuniform:
approaching \(r=r_+\) at fixed \(0<\theta<\pi\) gives the regular
nonpolar horizon, whereas approaching either axis retains the
divergence in Eq.~\eqref{eq:K-horizon-pole-divergence}.

Interestingly, the extremal case \(r_+=r_-\) implies that the horizon is free of the axial divergence.

\section{Schematic radial causal structure}
\label{app:causal-structure}

Here we shall summarize the radial causal structure of the
nonextremal horizon-selected branch \(q^2=\rplus^2\). The
discussion refers to a fixed nonaxial angle
\(0<\theta<\pi\), or equivalently to a nonpolar domain
\(\mathcal D_\epsilon\). It therefore suppresses the singular
rotation axes.

The surface \(r=\rplus\) is a regular Killing horizon in the
nonpolar region. Horizon-penetrating coordinates extend the metric
smoothly across it, separating the exterior region
\(r>\rplus\) from the black-hole interior
\(\rminus<r<\rplus\).

The causal structure differs from Kerr at the inner root. On the
horizon-selected branch, the curvature behaves as
\begin{equation}
  \Kretsch
  =
  \frac{C_-(\theta)}{r-\rminus}
  +
  \Ord(1) \,,
  \qquad
  r\rightarrow\rminus \,,
  \label{eq:causal-inner-divergence}
\end{equation}
where \(C_-(\theta)\) is generically nonzero at fixed
\(0<\theta<\pi\). Thus \(r=\rminus\) is a null curvature
singularity rather than a regular Cauchy horizon.

The connected extension from the exterior therefore terminates at
the inner null singularity. Unlike maximally extended Kerr, it
does not continue through a regular inner horizon into additional
asymptotic regions. In particular, the familiar infinite Kerr
tower of exterior regions is absent.

\begin{figure}[t]
  \centering
  \includegraphics[
    width=0.55\columnwidth
  ]{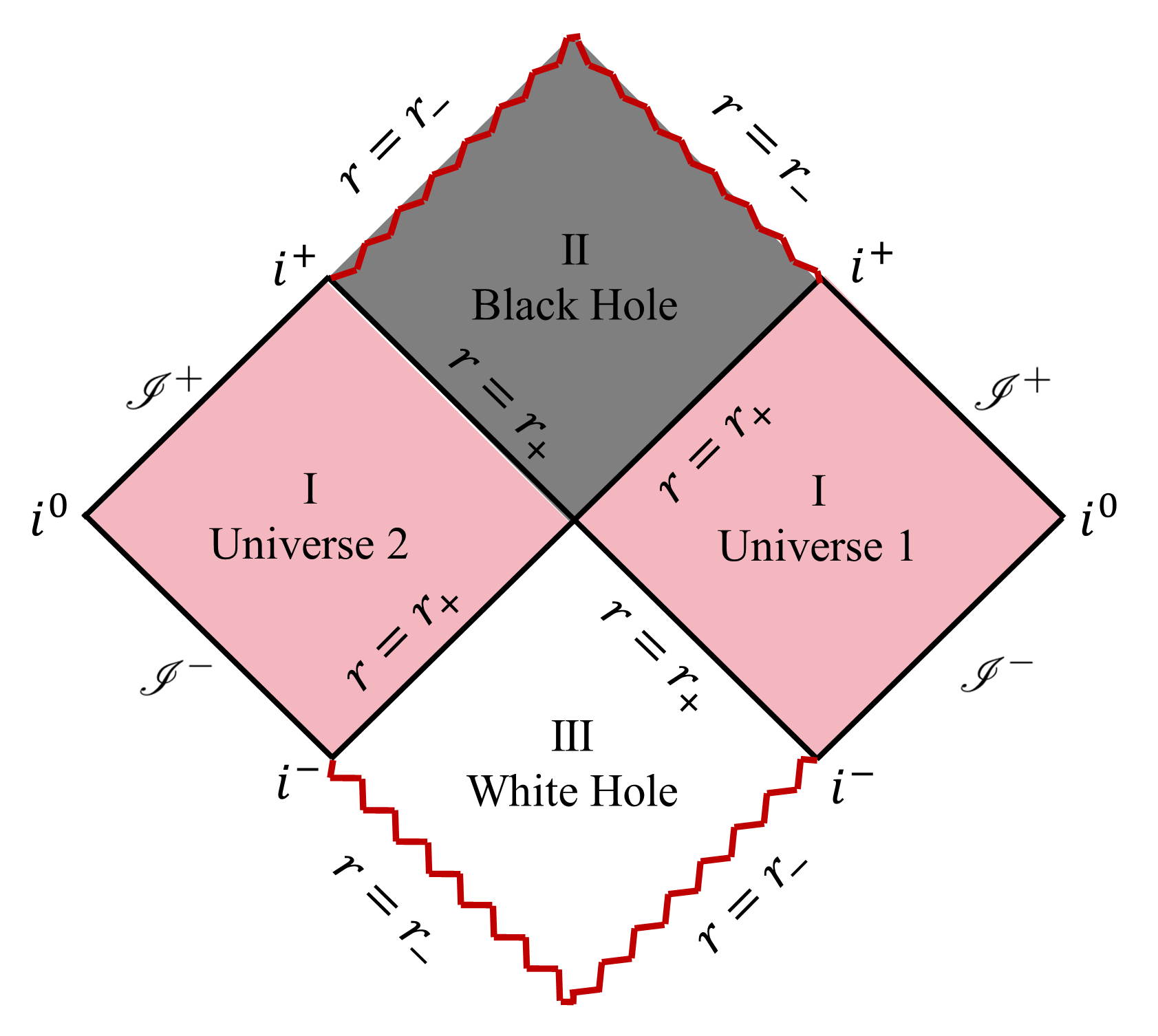}
  \caption{
    Schematic radial causal structure of the nonextremal
    horizon-selected branch \(q^2=\rplus^2\), at fixed nonaxial
    angle. The outer surface \(r=\rplus\) is a regular Killing
    horizon, whereas \(r=\rminus\) is a null curvature
    singularity. The axial curvature singularities at
    \(\theta=0,\pi\) are suppressed and are not represented in
    this two-dimensional diagram.
  }
  \label{fig:radial-causal-structure}
\end{figure}

Figure~\ref{fig:radial-causal-structure} gives a schematic
representation. The outer diagonal lines denote the regular
future and past branches of \(r=\rplus\), while the inner wavy
null boundaries represent the singular surface \(r=\rminus\).

Several qualifications are essential. First,
Fig.~\ref{fig:radial-causal-structure} represents only the
\((t,r)\) causal structure of the regular nonpolar sector. The
full four-dimensional spacetime contains two curvature-singular
axes that extend through the exterior and intersect the outer
horizon at its poles.

Second, the diagram does not establish a globally regular
bifurcate black-hole extension. The fixed-point obstruction shows
that a smooth symmetry-inheriting vector vacuum of strictly
nonzero norm cannot extend through the poles of a regular
nondegenerate bifurcation surface. Thus the bifurcation surface
shown in the reduced radial diagram should be understood as the
bifurcation structure of the punctured nonpolar sector, not as a
smooth compact two-surface of the complete spacetime.

Third, the diagram assumes that the curvature expansion
\eqref{eq:causal-inner-divergence} corresponds to a null singular
boundary in a horizon-penetrating chart. This should be verified
from the radial null characteristics or from the norm of the
\(r=\mathrm{const}\) hypersurfaces.

\section{Closedness and the separable obstruction}
\label{app:separable-obstruction}

For a closed stationary and axisymmetric one-form, Cartan's
identity gives
\[
  \Lie_X\Omega
  =
  \dd(\iota_X\Omega)+\iota_X\dd\Omega \,.
\]
Here \(\iota_X\Omega=\Omega(X)\) is the interior product.
Stationarity, axisymmetry, and \(\dd\Omega=0\) imply that
\(\iota_k\Omega=-E\) and \(\iota_m\Omega=L\) are constant on a
connected domain. 

Consider $\Omega=\Omega_r dr+\Omega_\theta d\theta$. Locally, \(\Omega=\dd S\). It follows that the norm condition of the bumblebee field $B^\mu B_\mu=b_0^2$ reduces to  a Hamilton-Jacobi (HJ) equation
\be \Delta(r)\Big(\frac{\partial S}{\partial r} \Big)^2+\Big( \frac{\partial S}{\partial\theta}\Big)^2=\Sigma(r\,,\theta) \,,\label{HJ-equation}\ee 
where \(\Sigma=r^2+a^2\cos^2\theta\) and \(\Delta=r^2-2Mr+a^2\). 

\subsection{Separable expansion}

Firstly, let us consider a more general separable solution
\[
  S=-Et+L\phi+S_r(r)+S_\theta(\theta) \,.
\]
Axis smoothness requires \(L=0\). The constant positive norm
then gives
\begin{align}
  &\Delta(S_r')^2+(S_\theta')^2
  -
  \frac{E^2(r^2+a^2)^2}{\Delta}
  +
  a^2E^2\sin^2\theta
  \nonumber\\
  &\hspace{24mm}
  = r^2+a^2\cos^2\theta \,.
  \label{eq:HJ-equation}
\end{align}
Separating the radial and angular terms gives
\begin{equation}
  (S_\theta')^2
  +
  a^2E^2\sin^2\theta
  -
 a^2\cos^2\theta
  =
  \lambda \,.
  \label{eq:HJ-angular}
\end{equation}

Smoothness requires \(S_\theta'=\Ord(\sin\theta)\). Evaluating
Eq.~\eqref{eq:HJ-angular} at the axis fixes
\(\lambda=-a^2\), and hence
\[
  (S_\theta')^2
  =
  -a^2(1+E^2)\sin^2\theta.
\]
No real axis-smooth separated solution exists for \(a\neq0\).

For \(E=0\), and a nonnegative separation constant
\(q^2\), the formal separated solution becomes Eq.~\eqref{eq:Kerr-oneform}.  Explicitly, one has
\be S=\int^r \sqrt{\frac{r'^2-q^2}{\Delta(r')}}\,\dd r'+\int^\theta \sqrt{a^2\cos^2\theta'+q^2}\,\dd\theta' \,,\label{solution-S}\ee
and
\begin{equation}
  \Omega
  =
  \sqrt{\frac{r^2-q^2}{\Delta}}\,\dd r
  +
  \sqrt{a^2\cos^2\theta+q^2}\,\dd\theta \,.
  \label{eq:Kerr-oneform-1}
\end{equation}
The nonzero polar limit of $\Omega$ is precisely
the axial failure identified in the main text.

\subsection{General real-envelope solutions}
\label{app:general-envelope-solutions}

The separated complete integral can be used to generate a more
general local envelope. Let
\begin{equation}
S_F(r,\theta;Q)
=
S(r,\theta;Q)+F(Q),
\qquad
Q\equiv q^2,
\label{eq:general-envelope-functional}
\end{equation}
where \(F\) is arbitrary. The envelope parameter
\(Q=Q(r,\theta)\) is determined implicitly by
\begin{equation}
\frac{\partial S_F}{\partial Q}=0.
\label{eq:general-envelope-condition}
\end{equation}
For the complete integral used here, this condition reads
\begin{align}
0
={}&
-\frac12
\int^r
\frac{\dd r'}
{
\sqrt{\Delta(r')}
\sqrt{r'^2-Q}
}
+
\frac12
\int^\theta
\frac{\dd\theta'}
{
\sqrt{a^2\cos^2\theta'+Q}
}
+
F'(Q).
\label{eq:explicit-envelope-condition}
\end{align}
The unspecified lower limits in the two integrals contribute only
a function of \(Q\) and can therefore be absorbed into \(F(Q)\).

On the envelope,
\begin{align}
\dd S_F
={}&
\left.
\partial_rS_F
\right|_Q\dd r
+
\left.
\partial_\theta S_F
\right|_Q\dd\theta+
\frac{\partial S_F}{\partial Q}\dd Q.
\end{align}
The last term vanishes by
Eq.~\eqref{eq:general-envelope-condition}. Hence
\begin{equation}
\Omega_r^2
=
\frac{r^2-Q(r,\theta)}{\Delta},
\qquad
\Omega_\theta^2
=
a^2\cos^2\theta+Q(r,\theta).
\label{eq:general-envelope-components}
\end{equation}

We restrict attention to envelopes generated by complete integrals
that are real on the full exterior domain
\begin{equation}
r\geq r_+,
\qquad
0\leq\theta\leq\pi.
\end{equation}
Reality of the angular integral at the equator requires
\(Q\geq0\). Reality of the radial integral for every
\(r'\geq r_+\) requires
\begin{equation}
r'^2-Q\geq0
\qquad
\text{for all }r'\geq r_+,
\end{equation}
and hence
\begin{equation}
0\leq Q\leq r_+^2.
\label{eq:global-real-envelope-bound}
\end{equation}
Accordingly, every envelope generated by this globally real
complete-integral family satisfies
\begin{equation}
0\leq Q(r,\theta)\leq r_+^2.
\end{equation}

At either rotation axis,
Eq.~\eqref{eq:general-envelope-components} gives
\begin{equation}
\Omega_\theta^2\big|_{\theta=0,\pi}
=
a^2+Q_{\rm ax}(r)>0,
\qquad
Q_{\rm ax}(r)\equiv Q(r,0),
\label{eq:general-envelope-axis-obstruction}
\end{equation}
for every rotating real envelope. Thus allowing the separation
parameter to vary does not restore one-form smoothness.

The axial conicity of the corresponding disformal metric is
\begin{equation}
\alpha^2(r)
=
1-
\frac{
\ell[a^2+Q_{\rm ax}(r)]
}{
(1+\ell)(r^2+a^2)
}.
\label{eq:general-envelope-conicity}
\end{equation}
Constant conicity on a connected exterior axial segment requires
\begin{equation}
a^2+Q_{\rm ax}(r)
=
C(r^2+a^2),
\label{eq:constant-conicity-profile}
\end{equation}
where \(C\) is constant.

Regularity of the nonpolar outer horizon requires
\begin{equation}
Q(r_+,\theta)=r_+^2,
\label{eq:general-envelope-horizon-condition}
\end{equation}
and in particular
\(Q_{\rm ax}(r_+)=r_+^2\).
If the constant-conicity profile extends continuously to the outer
horizon, Eq.~\eqref{eq:constant-conicity-profile} then fixes
\(C=1\), so that
\begin{equation}
Q_{\rm ax}(r)=r^2.
\label{eq:special-envelope-axis-profile}
\end{equation}
For every \(r>r_+\), however,
\begin{equation}
Q_{\rm ax}(r)=r^2>r_+^2,
\end{equation}
which contradicts the global-reality bound
\eqref{eq:global-real-envelope-bound}. Thus the special
constant-conicity profile cannot belong to a horizon-compatible
envelope generated by complete integrals that are real throughout
the exterior.

A locally real near-axis solution with
\(Q_{\rm ax}=r^2\) may still solve the envelope integrability
condition in a restricted neighborhood. Such a local branch does
not define a member of the globally real exterior family considered
here and therefore does not evade the global axial obstruction.

It follows that a globally real, horizon-compatible envelope cannot
make the conicity constant along the complete exterior axis.
Consequently, the exterior axis contains points at which
\begin{equation}
\partial_z\alpha\neq0.
\end{equation}
At every such point, the varying-conicity criterion gives
\begin{equation}
R_{\hat\rho\hat\phi\hat z\hat\phi}
=
-\frac{\partial_z\ln\alpha}{\rho}
+
\mathcal O(1),
\end{equation}
and excludes a \(C^2\) axial extension.

The complete scalar-invariant expansion of a nonseparable envelope
also contains derivatives of \(Q(r,\theta)\). The conclusion above
therefore establishes the unavoidable varying-conicity
frame-curvature obstruction somewhere on the complete exterior
axis; the constant-\(q\) family analyzed in the main text admits the
stronger direct result
\(\mathcal K\sim\rho^{-2}\) on every generic open axial segment..

\end{document}